\documentclass[letterpaper]{article} 
\usepackage{aaai25}  
\usepackage{times}  
\usepackage{helvet}  
\usepackage{courier}  
\usepackage[hyphens]{url}  
\usepackage{graphicx} 
\usepackage{natbib}  
\usepackage{caption} 
\usepackage{xcolor}%
\usepackage{algorithm}
\usepackage{algorithmic}
\usepackage{amsmath}
\usepackage{booktabs}
\usepackage{dsfont} 
\usepackage{multirow} 

\renewcommand{\eqref}[1]{\ref{#1}}

\newcommand{\cutsectionup}{\vspace*{-0.03in}}
\newcommand{\cutsubsectionup}{\vspace*{-0.02in}}
\newcommand{\cutparagraphup}{\vspace*{-0.05in}}
\newcommand{\cutcaptionup}{\vspace*{-0.065in}}

\title{Learning Multidimensional Urban Poverty Representation with Satellite Imagery}

\author{Sungwon Park\textsuperscript{\rm 1,2}, Sumin Lee\textsuperscript{\rm 1,2}, Jihee Kim\textsuperscript{\rm 2}, \\ Jae-Gil Lee\textsuperscript{\rm 2}, Meeyoung Cha\textsuperscript{\rm 1,2}, Jeasurk Yang\textsuperscript{\rm 1}, Donghyun Ahn\textsuperscript{\rm 1}}
\affiliations{
    \textsuperscript{\rm 1}   Max Planck Institute for Security and Privacy (MPI-SP) \\
    \textsuperscript{\rm 2}  Korea Advanced Institute of Science and Technology (KAIST)   
}

\usepackage{bibentry}

\begin{document}
\maketitle
\begin{abstract}
Recent advances in deep learning have enabled the inference of urban socioeconomic characteristics from satellite imagery. However, models relying solely on urbanization traits often show weak correlations with poverty indicators, as unplanned urban growth can obscure economic disparities and spatial inequalities. To address this limitation, we introduce a novel representation learning framework that captures multidimensional deprivation-related traits from very high-resolution satellite imagery for precise urban poverty mapping. Our approach integrates three complementary traits: (1) accessibility traits, learned via contrastive learning to encode proximity to essential infrastructure; (2) morphological traits, derived from building footprints to reflect housing conditions in informal settlements; and (3) economic traits, inferred from nightlight intensity as a proxy for economic activity. To mitigate spurious correlations—such as those from non-residential nightlight sources that misrepresent poverty conditions—we incorporate a backdoor adjustment mechanism that leverages morphological traits during training of the economic module. By fusing these complementary features into a unified representation, our framework captures the complex nature of poverty, which often diverges from economic development trends. Evaluations across three capital cities—Cape Town, Dhaka, and Phnom Penh—show that our model significantly outperforms existing baselines, offering a robust tool for poverty mapping and policy support in data-scarce regions.

\end{abstract}
\cutsectionup
\section{Introduction}
Measuring local socioeconomic conditions at a fine-grained level is a critical challenge for policymakers, traditionally reliant on census surveys that are costly and conducted infrequently. To overcome these limitations, recent work has increasingly turned to deep learning models that analyze widely available satellite imagery~\cite{jean2016combining, han2020lightweight, park2022learning, yan2024urbanclip}. A foundational method involves using nightlight intensity as a proxy signal to train models on high-resolution daytime imagery for socioeconomic estimation~\cite{jean2016combining}. Building on this, other approaches estimate the scale of economic development~\cite{han2020lightweight} or predict economic output like regional GDP from daytime imagery~\cite{yan2024urbanclip}.

\begin{figure}[t!]
\centering
\includegraphics[width=0.95\linewidth]{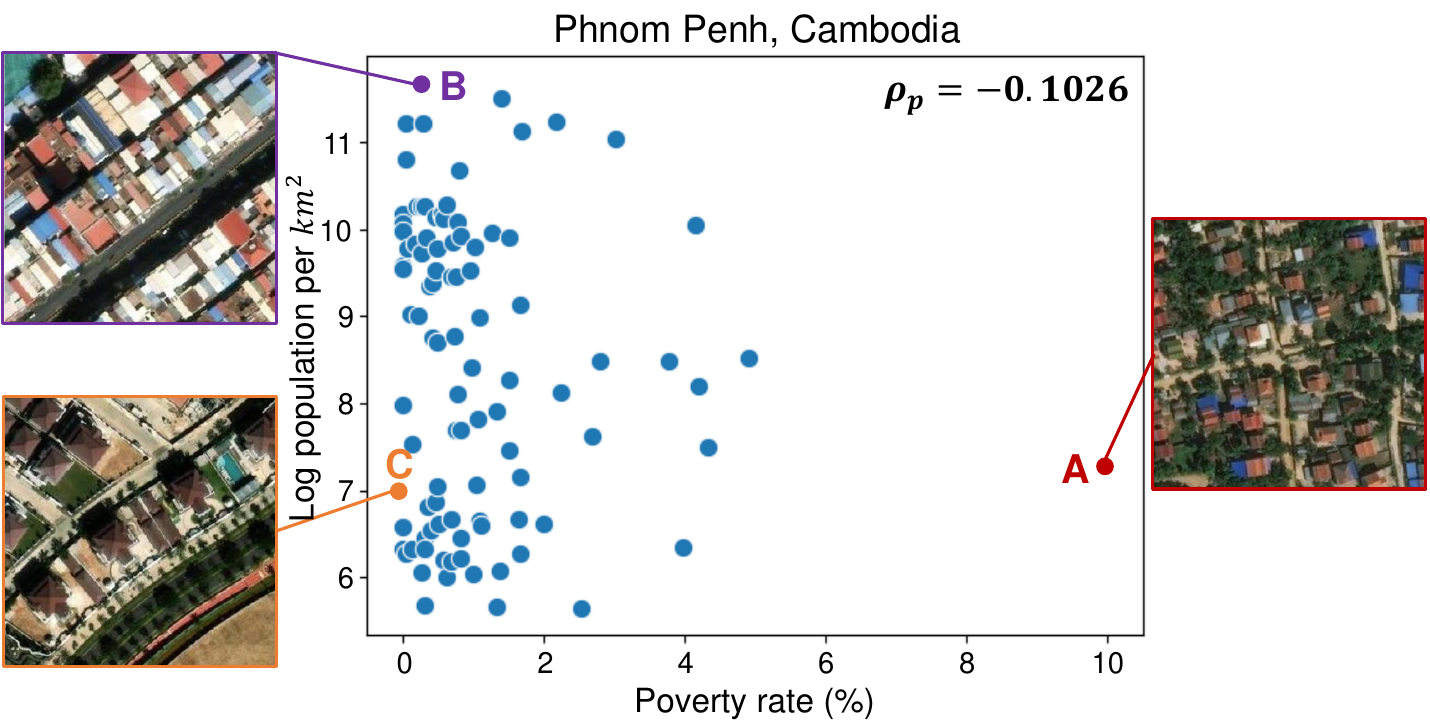}
\caption{Population density and poverty rates by district in Phnom Penh, Cambodia. Poverty does not consistently correlate with population density, based on the Pearson correlation ($\rho_p$). District B exhibits a more pronounced urbanization pattern with higher population density and compact land use than District C, yet their poverty rates remain similar. In contrast, District A and District C have comparable levels of urbanization, but District A experiences a significantly higher poverty rate (A: Kantaok, B: Ou Ruessei Ti Bei, C: Veal Sbov).}
\label{fig:poverty_intro}
\end{figure}

When these techniques are applied to the specific challenge of poverty mapping, however, their focus on proxies for economic growth and urbanization proves insufficient. Unplanned urban growth can produce dense informal settlements where high population density and compact land use—typically signs of development—mask the realities of deep poverty and inadequate housing~\cite{kraff2020morphologic, Kuffer2016}. Conversely, economic development signals can be actively misleading; bright nightlights from industrial zones or major infrastructure can create an illusion of economic vibrancy in areas with few residents, confounding poverty estimates. Our data from Phnom Penh, Cambodia, in Figure~\ref{fig:poverty_intro} support this. There is no consistent correlation between population density and poverty rates: districts with vastly different population densities can share similar poverty levels, while others with comparable densities exhibit a wide range of poverty rates. This demonstrates the unreliability of conventional urbanization metrics for estimating poverty.

This divergence occurs because poverty is a multidimensional phenomenon, a concept that cannot be captured solely by monetary or development proxies~\cite{UNHABITAT2022report, OPHImulti}. This perspective is formalized in frameworks like the Multidimensional Poverty Index (MPI)~\cite{UNDPmulti, OPHImulti}, which posits that a comprehensive understanding of poverty requires assessing deprivations across interconnected economic, environmental, and social domains. These include factors like poor housing conditions~\cite{kraff2020morphologic, Kuffer2016}, a lack of access to essential services~\cite{aguilar2020poverty}, and limited economic opportunities~\cite{UNDPmulti}, which are often defining characteristics of impoverished areas.

To address the challenge of modeling the multidimensional aspects of poverty, our framework is designed to learn representations that directly capture the socio-environmental signatures of deprivation from satellite imagery. We move beyond simple proxies by integrating three complementary traits:
\begin{itemize}
    \item \textbf{Accessibility Trait:} To quantify access to essential services—a key social dimension of poverty—we learn a representation that encodes a location's proximity to infrastructure such as schools, hospitals, and banks. The modeling is inspired by contrastive learning methods~\cite{radford2021learning}.
    \item \textbf{Morphological Trait:} To capture environmental conditions and living standards, this trait is derived from building footprints to reflect housing quality and the distinct structural patterns of informal settlements~\cite{Kuffer2016, kraff2020morphologic}.
    \item \textbf{Economic Trait:} To capture local economic activity, this trait is inferred from nightlight intensity.
\end{itemize}
These traits are then fused into a unified representation. Crucially, to make our economic trait robust and resolve the issue of confounding nightlight signals from non-residential areas, we incorporate a backdoor adjustment mechanism. This technique leverages our morphological trait module to identify and mitigate the influence of spurious light sources during training, creating a more robust economic indicator.

Evaluations conducted in urban areas across three low- and middle-income countries (LMICs)—Cape Town, Dhaka, and Phnom Penh—demonstrate that our approach significantly outperforms existing baseline models in poverty mapping. By integrating these carefully designed economic, morphological, and accessibility features, our method surpasses conventional urbanization-based proxies, allowing for a more accurate assessment of the complexities of poverty. This enhanced representation learning framework offers significant potential to inform data-driven policy-making and resource allocation, particularly in regions where traditional poverty measurement methods remain limited.

\cutsectionup
\section{Related Work}

\cutsubsectionup
\subsection{Representation Learning for Satellite Imagery}
In recent years, satellite imagery has become a widely used tool for socioeconomic predictions, providing a more efficient alternative to traditional resource intensive survey methods~\cite{han2020learning,ahn2023human,han2024geosee}. Representations are typically learned from satellite imagery at the grid level, which are then used for downstream tasks such as predicting socioeconomic indicators at the district level. Existing research can be categorized into two main approaches: the \textit{Proxy supervised approach} and the \textit{Contrastive learning approach}.
\cutparagraphup
\paragraph{Proxy supervised approach.}

The novelty of the proxy supervised approach lies in how it constructs a proxy label dataset that is strongly correlated with the target variable. An example is Jean et al. which used nightlight intensity as a proxy label for satellite imagery to extract representations associated with the relative wealth index in five African countries~\cite{jean2016combining}. Furthermore, \textsf{READ} introduces a weakly supervised method by creating a small set of human-labeled data to measure various socioeconomic variables at the district level~\cite{han2020lightweight}.

\cutparagraphup
\paragraph{Contrastive learning approach.}The contrastive learning approach emphasizes the construction of positive and negative samples to derive meaningful representations, , drawing inspiration from self-supervised learning.~\cite{park2021improving,han2020mitigating}. This method is initially based on Tobler's First Law of Geography, which states ``everything is related to everything else, but near things are more related than distant things." For instance, \textsf{Tile2Vec} applies this principle by creating triplet sets for training, where geographically adjacent satellite images are considered positive samples, and all other images are treated as negative samples~\cite{jean2019tile2vec}. On the other hand, \textsf{PG-SimCLR} suggests an alternative approach to the construction of positive samples using points of interest (POI)~\cite{xi2022beyond}. It demonstrates that the geographic distributions of socioeconomic indicators do not always adhere to Tobler's First Law, highlighting the need for a different strategy in contrastive learning. Recently, UrbanCLIP has utilized text descriptions of satellite imagery from Vision Language Models (VLM) as positive samples for CLIP loss, with the aim of maximizing the multimodal potential of satellite imagery~\cite{yan2024urbanclip}.

\cutsubsectionup
\subsection{Measuring Multidimensional Poverty with Satellite Imagery} 
Traditional poverty measurement relied on income or consumption expenditures per capita, with poverty lines defining thresholds below which individuals or households were considered poor (e.g., \$2.15 per day at 2017 purchasing power parities, PPP, for the World Bank's international poverty line). However, such an approach faced significant limitations, as monetary constraints alone do not fully capture the complex and interrelated deprivations that define poverty~\cite{Bourguignon2003multi}.

\begin{figure*}[t!]
\centering
\includegraphics[width=0.85\textwidth]{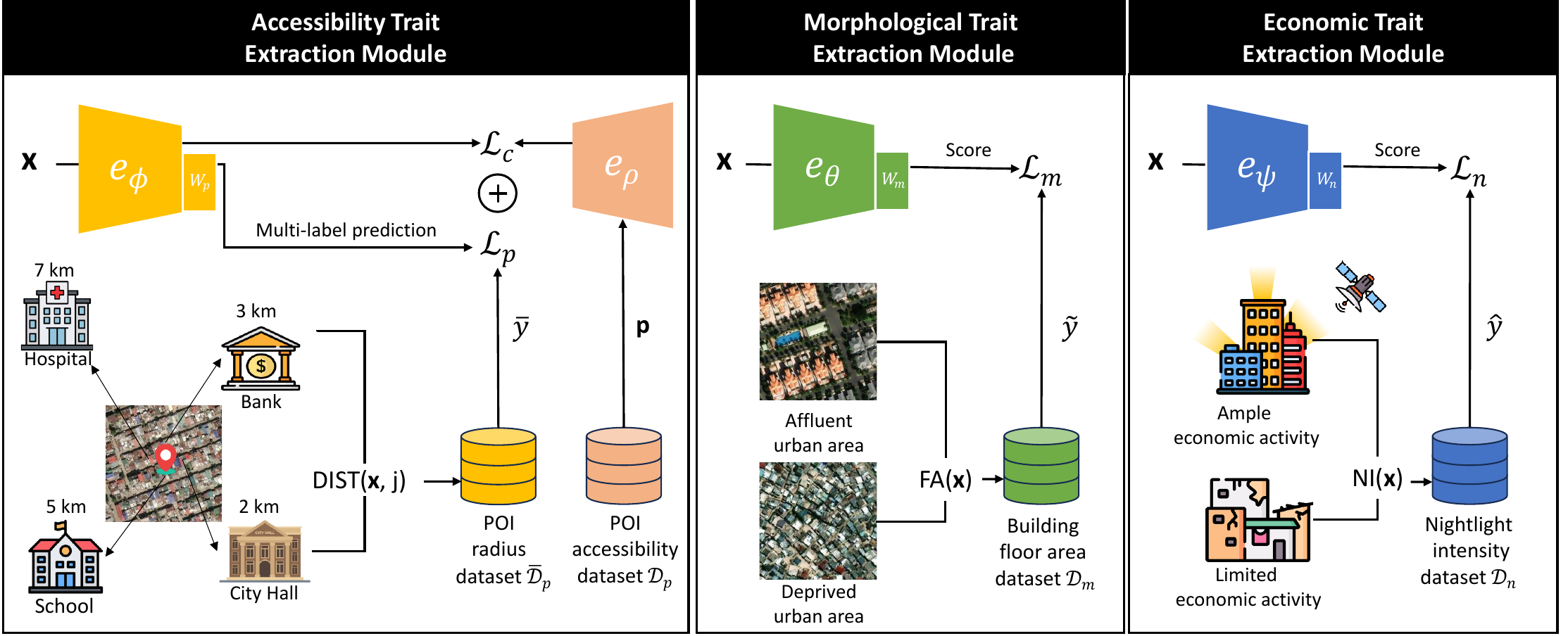}
\cutcaptionup
\caption{Illustration of the model architecture. The proposed framework has three main components: (1) the Accessibility Trait ; (2) the Morphological Trait; and (3)
the Economic Trait Extraction
Modules.}
\label{fig:main_model}
\end{figure*}

Alternative approaches have increasingly embraced multidimensional poverty measurement. Those approaches are theoretically rooted in Amartya Sen’s capability approach~\cite{Sen1999}, which emphasizes that poverty should be understood as deprivation in capabilities—such as education, health, and living standards—rather than merely as a lack of income. This shift was institutionalized in global policy frameworks with the introduction of the Multidimensional Poverty Index (MPI) by the United Nations Development Programme~\cite{UNDPmulti} and the Oxford Poverty and Human Development Initiative (OPHI)~\cite{OPHImulti}. For example, the MPI assesses poverty across three dimensions: health (e.g., child mortality, nutrition), education (e.g., years of schooling, school attendance), and living standards (e.g., access to electricity) (See Table~5 in the Appendix for details). This evolution in poverty measurement has important implications of identifying overlapped deprivations within households and provide a more accurate understanding of poverty’s structural determinants.

In the context of satellite imagery and deep learning applications, the shift toward multidimensional poverty measurement has also encouraged scholars to capture a broader range of socioeconomic indicators. Early studies~\cite{jean2016combining} focused on predicting consumption levels from satellite imagery. Subsequent research expanded this approach by estimating non-monetary indicators, such as access to electricity~\cite{Ratledge2022Nature}, thereby providing a more comprehensive understanding of deprivation. However, current approaches have not yet systematically categorized the features extracted from satellite imagery according to internationally recognized poverty dimensions. Instead, the studies typically generate a single feature representation from satellite images to estimate poverty levels, limiting alignment with established multidimensional frameworks.

In contrast, our proposed method employs a multidimensional approach that integrates economic, morphological, and accessibility traits, providing a more comprehensive representation of poverty. This tripartite classification is adopted because these three dimensions encapsulate the most informative aspects of remote sensing data for poverty assessment, offering a structured and interpretable framework for analyzing spatial deprivation.
\section{Method}
Assume an urban area composed of multiple districts with varying shapes and sizes. Let the set of satellite images for the $i$-th district be denoted as $\mathcal{D}_{i} = \{ \mathbf{x}_1, \dots, \mathbf{x}_{n_i}\}$, where $n_i$ represents the number of satellite images within the district. Our objective is to estimate the poverty indicator $y_i$ for $i$-th district, given the set of images $\mathcal{D}_{i}$. We propose a method designed to represent urban poverty through three distinct perspectives, as illustrated in Figure~\ref{fig:main_model}: (1)  Accessibility
traits, (2) Morphological traits, and (3)  Economic traits. If a model attempts to optimize for multiple tasks simultaneously, the shared representations may not ensure optimal performance due to the low correlation between tasks. To address this issue, we propose task-specific training using separate datasets and training signals, as outlined below:

\begin{itemize} 
    
    \item \textbf{Accessibility Trait Extraction Module} trains the encoder $e_\phi$ by explicitly aligning the image features with the corresponding distance information from each POI.

    \item \textbf{Morphological Trait Extraction Module} trains the encoder $e_\theta$ by maximizing the correlation between the model's predictions from satellite imagery and average building floor area size.

    \item \textbf{Economic Trait Extraction Module} trains the encoder $e_\psi$ by maximizing the correlation between the model's predictions from satellite imagery and nightlight intensity. To mitigate the effect of confounding variables, we apply backdoor adjustment by identifying semantically meaningful patches, guided by attention cues derived from morphological trait modules.
\end{itemize}

\cutsubsectionup
\subsection{Accessibility Trait Extraction Module}
\label{sec:accessibility}
 Essential infrastructure accessibility is a key determinant in poverty measurement, as it plays a fundamental role in meeting basic human needs. However, in low- and middle-income countries, the availability of infrastructure POIs is often limited, resulting in the frequent absence of corresponding POIs in individual satellite images.
Our key approach quantifies spatial relationships in satellite images by encoding the distance information to the closest POI, rather than directly using the POI present in the satellite images. Here, we train the encoder $e_\phi$ using a contrastive learning scheme by maximizing the similarity between the satellite image representation and the representation of the distance information from the POIs.
\cutparagraphup
\paragraph{Accessibility trait encoder training with POI distance information.} We begin by selecting essential infrastructure POIs, such as hospitals, schools, city halls, and banks, because their accessibility and availability directly impact key aspects of poverty, serving as foundational pillars for health, education, governance, and financial services. The distance vector $\mathbf{d}$ can be computed by measuring the distance from the center of the satellite imagery to each POI~(\eqref{eq:dist}). 

 However, the low-dimensional distance vector $\mathbf{d}$ lacks the capacity to capture rich high-frequency information from the satellite image~\cite{vivanco2024geoclip}. To address this limitation, we transform the distance vector $\mathbf{d}$ into a high-dimensional feature representation. Specifically, we denote $j$-th POI embedding vector as $\mathbf{p}_j$. Inspired by the gravity model, which describes an inverse relationship between poverty and distance in spatial interactions~\cite{anderson1979theoretical}, we perform element-wise weighted sum of these POI embeddings to obtain a joint representation $\mathbf{p}$. 
 
 Let the entire set of satellite imagery from entire urban area be denoted as $\mathcal{I}$. 
The dataset $\mathcal{D}_p$ for POI accessibility can be constructed by measuring the distance from the center of the satellite imagery to each POI, as follows:

\begin{align}
 \mathbf{d}^j &= \text{DIST}(\mathbf{x}, j) \label{eq:dist} \\ 
\mathcal{D}_p &= \{(\mathbf{x}, \mathbf{p}) \mid \mathbf{x} \in \mathcal{I} \text{ and } 
 \mathbf{p} = \sum_j \frac{\mathbf{p}_j}{\mathbf{d}^{j}}\}
\end{align}

\noindent where $\mathbf{d}^j$ is $j$-th element of distance vector $\mathbf{d}$ and $\text{DIST}(\mathbf{x}, j)$ represents the distance between the center of image $\mathbf{x}$ and the $j$-th POI.

Once the POI accessibility dataset $\mathcal{D}_p$ is constructed, the next step is to integrate the accessibility traits into the encoder $e_\phi$. 
Given the batch $\mathcal{X}$ from the accessibility dataset $\mathcal{D}_p$, the satellite and POI representations are aligned by maximizing the agreement between representations learned across different modalities while capturing different relationships from negative samples $\mathbf{p}^{n} \in \mathcal{X}$. This alignment is achieved through contrastive learning with the image encoder $e_\phi$ and the POI encoder $e_\rho$ as follows:

\begin{align}
   \mathcal{L}_{c} &= - \frac{1}{|\mathcal{D}_p|}\sum_{(\mathbf{x},\mathbf{p}) \in \mathcal{D}_p} \log \frac{\exp(\text{sim}(e_\phi(\mathbf{x}), e_\rho(\mathbf{p})) /\tau )}{\sum_{\mathbf{p}^{n}\in{\mathcal{X}}} \exp(\text{sim}(e_\phi(\mathbf{x}), e_\rho(\mathbf{p}^n))/\tau)}
   \label{eq:s_con}
\end{align}

\noindent where \text{sim} represents the cosine similarity metric between two vectors, and $\tau$ is the sharpening temperature.

Ensuring infrastructure is within a certain radius $\gamma$ allows residents to access essential services quickly and efficiently. To further enhance the learned representation, we introduce an additional design incorporating an accessibility prediction task. Given the distance vector $\mathbf{d}$ derived from the satellite image $\mathbf{x}$, the accessibility precondition loss $\mathcal{L}_p$ can be defined with the dataset multi-labels $\Bar{\mathcal{D}}_{p}$ as follows:

\begin{align}
\Bar{\mathcal{D}}_p &= \{(\mathbf{x}, \Bar{\mathbf{y}}) | \mathbf{x} \in \mathcal{X} \text{ and } \Bar{\mathbf{y}}^j = \mathds{1}(\mathbf{d}^j < \gamma) \}, \\
\mathcal{L}_{p} &= - \frac{1}{|\Bar{\mathcal{D}}_p|} \sum_{(\mathbf{x}, \bar{y}) \in \Bar{\mathcal{D}}_p}{H(\Bar{y}, W_p \cdot e_\phi(\mathbf{x}))}, \label{eq:loss_prediction} 
\end{align}
\noindent where $H$ represents the binary cross-entropy loss, and $W_p$ is a weight matrix that projects the encoder output into the multi-label space.

Finally, we train the accessibility trait 
 encoder $e_\phi$ by combining
the loss terms discussed earlier, with hyperparameters $\lambda_1$ as given in \eqref{eq:total_loss}. 

\begin{align}
    \mathcal{L}_{a} = \mathcal{L}_{c} + \lambda_{1}\mathcal{L}_p
    \label{eq:total_loss}
\end{align}

\cutsubsectionup
\subsection{Morphological Trait Extraction Module}
\label{sec:morphological}
While accessibility traits show promise in poverty estimation, relying solely on them fails to capture the visual manifestations of poverty.  To address this limitation, it is essential to directly extract the morphological traits of urban poverty from satellite imagery. Inspired by remote sensing studies on poverty, which suggest that impoverished areas often exhibit smaller building sizes due to limited land availability and self-built structures~\cite{Kuffer2016,yang2024poverty}, we incorporate the building floor area as a representation of the visual traits related to poverty from the encoder $e_\theta$.
\paragraph{Morphological trait encoder training with building floor area.} The building floor area dataset $\mathcal{D}_m$ is constructed using satellite imagery from the entire urban area as follows:

\begin{align}
\mathcal{D}_m &= \{(\mathbf{x}, \Tilde{y}) | \mathbf{x} \in \mathcal{X} \text{ and } \Tilde{y} = \text{FA}(\mathbf{x})\},
\end{align}

\noindent where FA($\mathbf{x}$) indicates the average building floor area, calculated as the total building area divided by the number of buildings corresponding to $\mathbf{x}$.

The morphological trait encoder $e_\theta$ is trained by maximizing the correlation between its predictions from satellite imagery and the corresponding floor area. The loss function is defined as:

\begin{align}
\mathcal{L}_{m} &= - \frac{1}{|\mathcal{D}_m|} \sum_{(\mathbf{x}, \tilde{y}) \in \mathcal{D}_m}{R(\tilde{y}, W_m \cdot e_\theta(\mathbf{x}))}, \label{eq:loss_behavioral} 
\end{align}

\noindent where $W_m$ is a weight matrix projecting the encoder output to the score, and $R$ is the Pearson correlation function.

\cutsubsectionup
\subsection{Economic Trait Extraction Module}
\label{sec:behavioral}

Nightlight intensity serves as a widely used proxy for economic traits of poverty, as low illumination typically reflects limited economic activity or resource constraints—key indicators of impoverished regions ~\cite{jean2016combining,yeh2020using}. However, existing methods that rely on this proxy may be influenced by confounding factors, leading the model to incorporate regions that do not accurately reflect true poverty conditions during training~\cite{wang2024nuwadynamics,hao2025nature}. This can introduce noise into the learning process and reduce the accuracy of the resulting representations. To address this, we extract the economic traits of poverty from the encoder $e_\psi$ by maximizing correlation with nightlight intensity while mitigating confounding effects through backdoor adjustment. Specifically, semantically salient patches—identified using attention cues from the morphological trait module—are selectively used during training to emphasize causal signals and reduce reliance on spurious correlations.
\cutparagraphup
 \paragraph{Economic trait encoder training with nightlight intensity.} 
 For a given input $\mathbf{x}$ $\in$ $\mathcal{I}$ , we perform a backdoor adjustment by replacing non-causal patches with semantic blocks from the causal patches. The process utilizes attention scores from the encoder $e_\theta$ and is formulated as follows:

\begin{align}
\hat{P_{i}} = 
\begin{cases} 
    P_i, & \text{if } A_{e_\theta}^{i} \geq \tau \text{ (causal)} \\
    \frac{1}{|\mathcal{N}(P_i)|} \sum_{P_j \in \mathcal{N}(P_i)} P_j, & \text{if } A_{e_\theta}^{i} < \tau \text{ (non-causal)}
\end{cases}
\end{align}

\noindent where $ P_i $ denotes the  $i$-th patch of image \( \mathbf{x} \), and \( \mathcal{N}(P_i) \) represents the set of neighboring causal patches. \( A_{e_\theta}^{i} \) is the attention value of patch \( P_i \) as determined by the encoder \( e_\theta \), and \( \tau \) is the threshold used to distinguish causal from non-causal patches.

 We created a nightlight intensity dataset 
$ \mathcal{D}_n$ for each satellite image $\mathbf{x}$ as follows: \begin{align}
\mathcal{D}_n &= \{(\hat{\mathbf{x}}, \hat{y}) | \mathbf{x} \in \mathcal{I} \text{ and } \hat{y} = \text{NI}(\mathbf{x})\},
\end{align}

\noindent where \( \hat{\mathbf{x}} \) is the backdoor-adjusted version of \( \mathbf{x} \), and \( \text{NI}(\mathbf{x}) \) denotes the average nightlight intensity associated with image \( \mathbf{x} \).

We train the economic trait encoder $e_\psi$ by maximizing the correlation between the model predictions from satellite imagery and nightlight intensity. The loss function to train the encoder with $\mathcal{D}_n$ is defined as the following equation:
\begin{align}
\mathcal{L}_{n} &= - \frac{1}{|\mathcal{D}_n|} \sum_{(\mathbf{x}, \hat{y}) \in \mathcal{D}_n}{R(\hat{y}, W_n \cdot e_\psi(\mathbf{x}))}, \label{eq:loss_behavioral} 
\end{align}

\noindent where $W_n$ is a weight matrix that projectes the encoder output to the score.
\cutsubsectionup
\subsection{Estimating Poverty Indicators for Individual Districts} 

Since districts vary in shape and size, the number of available satellite images per district can differ significantly, resulting in fluctuations in the extracted feature set size. To ensure a consistent district-level representation, it is crucial to generate fixed-length representations from a variable number of images in each district.

The district feature vector $\mathbf{r}_i$ for the $i$-th district $\mathcal{D}i$ is obtained using three distinct trait encoders, as defined as follows: \begin{align}
\mathbf{r}_{i} &= [ \mu(e_\theta(\mathcal{D}_i)),
\mu(e_\phi(\mathcal{D}_i)),  \mu(e_\psi(\mathcal{D}_i))], \label{eq:disctrict_vector}
\end{align} 

\noindent where $\mu$ represents the mean operation, and [·] denotes concatenation.

The corresponding district-level poverty indicator $y_i$ is then estimated using the regression model $h\varphi$ (i.e., $y_i = h_\varphi(\mathbf{r}_{i})$)

\cutsectionup
\section{Training Setups}
\cutsubsectionup
\subsection{Dataset}
To effectively train our model for poverty estimation, we leverage a diverse set of multimodal data sources that capture both spatial and socioeconomic characteristics of urban areas. These include very-high-resolution (VHR) satellite imagery, nightlight intensity, building footprint information, and POIs. 
\cutparagraphup
\paragraph{Poverty Ground Truth.} The poverty headcount population, which represents the number of individuals falling below a specific income threshold, is used as the ground truth.
Poverty-related indicators in three countries were obtained from census data.
For Dhaka, we use the \textit{Poverty Maps of Bangladesh}, compiled by the Bangladesh Bureau of Statistics and the World Food Programme, to derive district-level poverty estimates. For Phnom Penh, we rely on the \textit{Subnational Poverty Rate Data}, published by the Ministry of Planning, which provides district-level poverty headcount rates for Cambodia. In Cape Town, ward-level monthly household income data from the \textit{National Census} is used to calculate the percentage of the population below the the South African government's poverty line (3,200 rand per month).
\cutparagraphup

\paragraph{Satellite Images.} 
We utilize VHR daytime satellite images of urban areas in three LMICs: Phnom Penh, Cambodia (KHM); Dhaka, Bangladesh (BGD); and Cape Town, South Africa (ZAF). These cities were selected due to their high poverty levels and the availability of district-level poverty census data to evaluate our predictions. These images are sourced from WorldView-2 and GeoEye via the World Imagery service, accessed through an ArcGIS Developer subscription. Each image has a resolution of 0.6 meters per pixel and a size of 256 × 256 pixels.



\begin{table*}[t!]
    \centering
    \small 
    \resizebox{0.8\textwidth}{!}{
    \renewcommand{\arraystretch}{1.2} 
    \setlength{\tabcolsep}{6pt} 
    \begin{tabular}{l | ccc ccc ccc}
        \toprule
        \multirow{2}{*}{Method} & \multicolumn{3}{c}{Phnom Penh, Cambodia} & \multicolumn{3}{c}{Dhaka, Bangladesh} & \multicolumn{3}{c}{Cape Town, South Africa} \\
        \cmidrule(lr){2-4} \cmidrule(lr){5-7} \cmidrule(lr){8-10}
        & Pearson & Spearman & R² & Pearson & Spearman & R² & Pearson & Spearman & R² \\
        \midrule
        Nightlight Proxy & 0.667 & 0.666 & 0.435 & 0.632 & \underline{0.581} & \underline{0.297} & \underline{0.589} & \underline{0.654} & \underline{0.589} \\
        Tile2Vec         & 0.655 & 0.673 & 0.417 & 0.465 & 0.373 & 0.188 & 0.510 & 0.620 & 0.494 \\
        READ             & \underline{0.677} & 0.675 & 0.440 & 0.477 & 0.359 & 0.208 & 0.542 & 0.639 & 0.526 \\
        PG-Simclr        & 0.675 & \underline{0.677} & \underline{0.445} & \underline{0.634} & 0.560 & 0.296 & 0.579 & 0.647 & 0.588 \\
        UrbanCLIP        & 0.662 & 0.653 & 0.429 & 0.629 & 0.524 & 0.290 & 0.547 & 0.627 & 0.535 \\
        \midrule
        Ours             & \textbf{0.719} & \textbf{0.744} & \textbf{0.489} & \textbf{0.670} & \textbf{0.592} & \textbf{0.324} & \textbf{0.609} & \textbf{0.667} & \textbf{0.608} \\
        \multicolumn{1}{l|}{\textit{Improvement (\%)}}
                         & +6.2 & +9.9 & +9.7 & +5.7 & +2.0 & +9.0 & +3.5 & +2.0 & +3.3 \\
        \bottomrule
    \end{tabular}
    }
    \cutcaptionup
    \caption{Evaluation results for poverty headcount population prediction across three LMIC cities: Phnom Penh (Cambodia), Dhaka (Bangladesh), and Cape Town (South Africa).}
    \label{tab:main_results}
\end{table*}

\begin{table}[t!]
    \centering
    \resizebox{0.9\linewidth}{!}{%
    \renewcommand{\arraystretch}{1.2} 

    \begin{tabular}{lccc}
        \toprule
        Method & Pearson & Spearman & R² \\
        \midrule
        Full Component & \textbf{0.7190} & \textbf{0.7437} & \textbf{0.4886}  \\
        w/o Accessibility Trait & 0.7044 & 0.7258 & 0.4733 \\
        w/o Morphological Trait  & 0.7055 & 0.7245 & 0.4723 \\
        w/o Economic Trait & 0.7074	& 0.7223 &	0.4754 \\
        w/o Backdoor Adjustment & 0.7174 & 0.7315 & 0.4798  \\
        \bottomrule
    \end{tabular}
    }
\cutcaptionup
        \caption{Ablation study results for different model components on the Cambodia dataset.}

            \label{tab:ablation}
        
\end{table}
\cutparagraphup
\paragraph{Nightlight Intensity.} 
Nightlight data was obtained from the Visible Infrared Imaging Radiometer Suite (VIIRS) dataset, released by the Earth Observation Group (EOG). VIIRS captures global nightlight emissions at a resolution of 15 arc-seconds. We utilized the data from the Annual VNL V2 dataset, which has been processed to remove noise from sunlit, moonlit, and biomass burning sources~\cite{VIIR}.
\cutparagraphup
\paragraph{Building Floor Area (FA).} The average 
building floor area was used as a proxy for morphological quality and was calculated by aggregating building polygons within a unit area. The building footprint data, including structural shapes, was obtained from OpenStreetMap (OSM), a widely used, crowd-sourced mapping platform. The latest building data was extracted from the Geofabrik Download Server, which provides regularly updated OpenStreetMap data.
\cutparagraphup
\paragraph{Points of Interest (POI).} 
POI data was used to evaluate accessibility to key urban amenities, as previous studies indicate a strong correlation between human activity and POI distribution~\cite{aguilar2020poverty}. We obtained location data for hospitals, schools, city halls, and banks from the Humanitarian OpenStreetMap Team (HOTOSM) dataset to ensure comprehensive coverage of socially significant infrastructure. These four types of POIs—health, education, public, and financial—were selected based on HOTOSM data standards.

\subsection{Implementation Details} 
All models are evaluated under uniform experimental conditions to ensure a fair comparison, maintaining consistency in the backbone network, optimizer, and learning rate. We employ the ViT-L/14 transformer architecture as the backbone for the image encoder. Training is conducted using the AdamW optimizer with a learning rate of 1e-4 and a batch size of 32. The learnable temperature parameter \( \tau \) for contrastive learning is initialized to 0.07. The infrastructure accessibility threshold \( \gamma \) is set to a 2 km radius, and the hyperparameter \( \lambda_1 \), which adjusts the loss term weight, is set to 0.1. For backdoor adjustment, the threshold $\tau$ was set such that the bottom 30\% of patches (based on attention scores) were replaced. For inference, the regression model $h\varphi$ is fit with ground-truth poverty indicators. A random forest regressor with 50 estimator trees is used. Implementation details for the baseline models can be found in the Appendix.

\cutsectionup
\section{Experiment}
We evaluate the poverty representation capability of our proposed model across diverse datasets from LMICs, comparing it with contemporary baselines. We then examine the impact of various model components on overall performance.

\begin{table}[t!]
    \centering
    \resizebox{0.8\linewidth}{!}{%
    \renewcommand{\arraystretch}{1.2} 

    \begin{tabular}{lccc}
        \toprule
        Method & Pearson & Spearman & R² \\
        \midrule
Full Component & \textbf{0.7190} & \textbf{0.7437} & \textbf{0.4886} \\
w/o Hospital & 0.7120 &	0.7367	& 0.4792\\
w/o Bank & 0.7124	& 0.7384	& 0.4780 \\
w/o School & 0.7120	& 0.7378	& 0.4790\\
w/o Townhall & 0.7107	&0.7324 & 0.4769\\
w/o POI Distance & 0.7043 & 0.7195 & 0.4718 \\
        \midrule
        radius $\gamma = 1$ & 0.7138	& 0.7411	& 0.4805 \\
        radius $\gamma = 2$ &\textbf{0.7190} & 0.7437 & \textbf{0.4886} \\
        radius $\gamma = 3$ & 0.7170	& \textbf{0.7440}	& 0.4839\\
        radius $\gamma = 4$ & 0.7151	& 0.7408	& 0.4815 \\
        \bottomrule
    \end{tabular}
    }
\cutcaptionup
        \caption{Component analysis of the impact of accessibility traits on
the Cambodia dataset.}

            \label{tab:poi_analysis}
        
\end{table}

\cutsubsectionup
\subsection{Performance Evaluation}
We evaluate the district feature vector $\mathbf{r}_i$ in relation to the poverty headcount population $y_i$, following principles established in satellite representation learning literature~\cite{han2020lightweight,yan2024urbanclip}. To ensure robustness, we randomly split the data into training and test sets using an 80\%-20\% ratio, repeating this process 50 times. We use R-squared values and correlation as evaluation metrics to assess how well our predictions approximate the ground truth. The R-squared value quantifies portion of variance in the dependent variable explained by the regression model, while correlation captures the overall trend and strength of the relationship, independent of scale differences. 

\begin{figure*}[t!]
\centerline{
\includegraphics[width=0.85\linewidth]{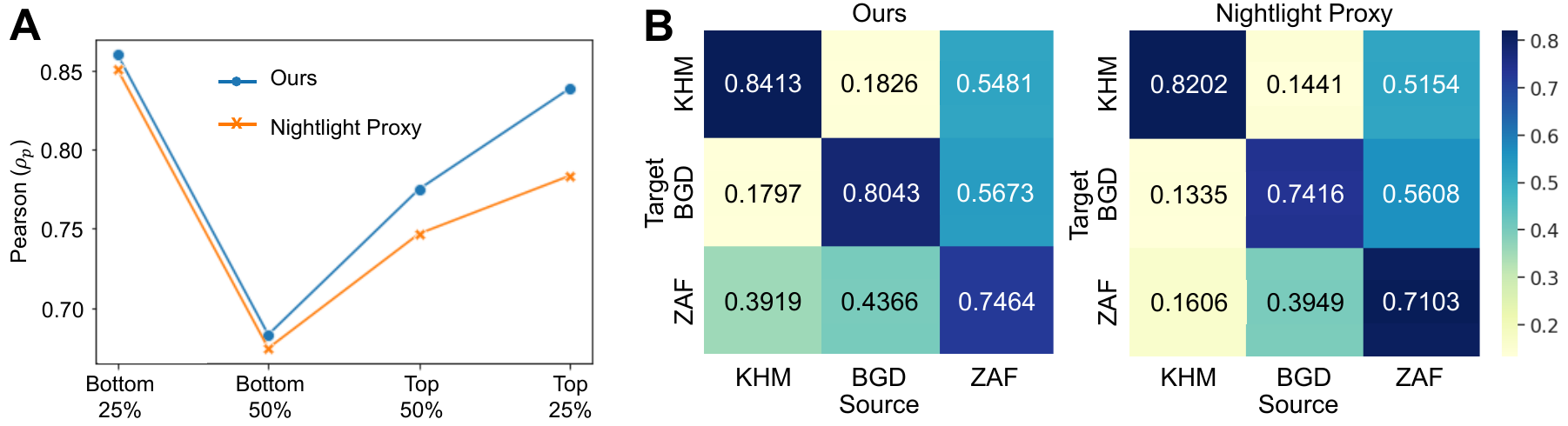}}
\cutcaptionup
      \caption{ Comparison of our model with the Nightlight Proxy. (A) Pearson correlation on the Cambodia dataset across districts grouped by ground-truth poverty headcount quartiles, (B) Pearson correlation heatmap for the transferability test (KHM: Cambodia, BGD: Bangladesh, ZAF: South Africa).}
\label{fig:nl_compare}
\end{figure*}


\cutparagraphup
\paragraph{Baselines.} We implemented several existing satellite representation learning models as baselines: (1) \textsf{Nightlight Proxy} utilizes nightlight intensity as a proxy for training the encoder. (2) \textsf{Tile2Vec} employs a triplet loss to learn satellite image representations in an unsupervised manner, using nearby images as positive samples and distant images as negative samples. (3) \textsf{READ} is a weakly supervised model that leverages a small set of human-labeled data, categorized into urban, rural, and uninhabited classes. (4) \textsf{PG-SimCLR} applies contrastive learning by constructing positive samples based on POI proximity. (5) \textsf{UrbanCLIP} utilizes multimodal learning by integrating satellite images with their corresponding text descriptions from a VLM model.

\cutparagraphup
\paragraph{Results.} Table~\ref{tab:main_results} demonstrates that our model outperforms all baseline methods across urban areas in three LMICs, consistently achieving superior performance on all metrics. In some instances, alternative baselines performed worse than the nightlight proxy baseline, underscoring that poverty does not necessarily correlate with urbanization. This highlights the importance of capturing the multidimensional nature of poverty.  All paired sample t-test p-values were well below 0.01, indicating statistically significant differences.

\cutparagraphup
\paragraph{Ablation Study.} 
To assess the contribution of each component, we conducted an ablation study by systematically removing individual components and comparing the following variations: (1) \textbf{Full Components}: The complete model, incorporating all components. (2) \textbf{w/o Accessibility Trait}: An ablation excluding the representation from the accessibility trait encoder \( e_\phi \). (3) \textbf{w/o Morphological Trait}: An ablation excluding the representation from the morphological trait encoder \( e_\theta \). (4) \textbf{w/o Economic Trait}: An ablation excluding the representation from the economic trait encoder \( e_\psi \).  (5) \textbf{w/o Backdoor Adjustment}: An ablation omitting the backdoor adjustment process. Table~\ref{tab:ablation} shows that the full model, incorporating all components, achieves the highest performance across all variations. This result underscores the importance of each component in capturing the multidimensional nature of poverty. 

\cutsectionup
\section{Discussion} 
\paragraph{Dissecting Urban Accessibility for Poverty Prediction.} 
In the ablation study with the Cambodia dataset, removal of POI distance representation leads to the most significant performance decline, underscoring its importance in poverty mapping. We go further by dissecting the impact of each type of POI, shown in Table~\ref{tab:poi_analysis}. The results show that accessibility to medical, financial, educational, government services are almost equally important in predicting poverty while the proximity to town hall has a slightly greater impact than the others. It suggests that considering multiple urban services collectively is important rather than focusing on a single urban service alone. Additionally, for the Cambodia's capital city, POI information within a two-kilometer radius seems the most relevant in predicting poverty, yields the highest predictive performance, indicating that this scale effectively captures the relationship between accessibility and poverty. Smaller (1 km) radius may be too localized, while larger (3-4 km) radius could dilute accessibility’s impact, as socioeconomic disparities tend to be concentrated in urban areas. These findings highlight the importance of optimizing urban accessibility in sensitivity analyses to enhance the accuracy of poverty mapping. Policymakers may find these insights useful for guiding infrastructure investments in impoverished areas, potentially improving access to essential services such as healthcare, education, financial institutions, and government support for marginalized communities.
\cutparagraphup
\paragraph{Improved Effectiveness and Transferability over the Nightlight Proxy.} 
We conducted a comparative evaluation between our model and the nightlight proxy model to assess the impact of incorporating multidimensional poverty traits. Figure~\ref{fig:nl_compare}(A) presents the evaluation results, where regions are categorized into poverty quartiles: bottom 25\%, bottom 50\%, top 25\%, and top 50\%. The results show our model's performance advantage is greatest in the most impoverished districts (the top quartiles). This suggests its strength lies in capturing the multidimensional aspects of severe poverty that are not reflected in purely economic indicators like nightlight intensity. Additionally, we evaluated the transferability of our model compared to the nightlight proxy, as shown in Figure~\ref{fig:nl_compare}(B). Our model demonstrates a 20.8\% improvement (0.318 $\rightarrow$ 0.384) over the nightlight proxy across six source-target pairs in three urban areas, highlighting its robustness and adaptability in diverse settings.

\cutsectionup
\section{Conclusion}
In this paper, we introduced a novel AI-driven approach to poverty analysis using satellite imagery, leveraging representation learning to model three key dimensions of poverty: accessibility, morphological, and  economic traits. 
Experiments across urban areas in three LMICs demonstrate that our model outperforms existing baselines in predicting poverty levels, highlighting its potential for data-driven policy-making and resource allocation. Overall, our work advances AI-driven poverty mapping by showing how established multidimensional poverty frameworks can directly inform the design of a representation learning model. Instead of relying on generalized proxies, we show how the specific dimensions outlined in multidimensional poverty frameworks, such as the MPI, can be used to engineer targeted features from satellite imagery. Our integration of accessibility, morphological, and economic traits is a direct implementation of this principle, translating abstract concepts of deprivation into concrete, measurable signals for a more nuanced and effective model. We hope our work contributes to sustainable development initiatives and encourages further research in AI-driven socioeconomic analysis.
\bibliography{aaai25}
\newpage
\newpage
\appendix
\section{APPENDIX}
\subsection{Further Details of Implementation}
\paragraph{Dataset.} 
We utilize VHR daytime satellite images of urban areas in three LMICs: Phnom Penh, Cambodia (KHM); Dhaka, Bangladesh (BGD); and Cape Town, South Africa (ZAF). The zoom level determines the spatial coverage of each tile, where a zoom level of $Z = k$ means that the tile spans $2^
k$ divisions of the Earth's longitude and latitude. At zoom level 0, a single tile covers the entire globe, and with each increment in 
$Z$, both the width and height of the tile are reduced by half. Since each tile maintains a fixed resolution of 256 × 256 pixels, increasing the zoom level enhances the spatial detail. For our experiments, we used images at zoom level 18, providing a resolution of 0.6 meters per pixel. The dataset consists of satellite images captured between 2022 and 2024. 
Table~\ref{tab:data_description} provides detailed statistics for each urban area. 

\begin{table}[h!]
    \centering

    \resizebox{0.7\linewidth}{!}{%
    \begin{tabular}{lcc}
        \toprule
        \textsf{City} & \textsf{\# of Districts} & \textsf{Total Images}   \\  
        \midrule
        Phnom Penh & 96 & 95,987 \\
        Dhaka & 41 & 226,305 \\
        Cape Town & 102 & 179,609 \\
        \bottomrule
    \end{tabular}
    }
    \caption{Dataset statistics for each city, including the number of districts, total images.}
    \label{tab:data_description}
\end{table}

The poverty headcount population, representing the number of individuals below a specific income threshold, is used as the ground truth. It is calculated by multiplying the district population by the corresponding poverty rate. he poverty headcount threshold is set at \$2.15 per month for Phnom Penh and Dhaka, while for Cape Town, it is defined as 3,200 rand per month. The average nightlight luminosity was computed based on the spatial extent of individual satellite images to account for the difference in zoom levels between nightlight satellite imagery and daytime satellite images.

\paragraph{Implementation details.} 
To ensure a fair comparison, all models are evaluated under consistent experimental conditions, maintaining uniformity in the backbone network, optimizer, and learning rate. We adopt the ViT-L/14 Transformer as the backbone for the image encoder. Training is performed using the AdamW optimizer with a learning rate of \(1e^{-4}\) and a batch size of 32. 4 V100 GPU is utilized to train. The learnable temperature parameter \( \tau \) for contrastive learning is initialized at 0.07. Additionally, the infrastructure accessibility threshold \( \gamma \) is set to 2 km, and the loss term weight hyperparameter \( \lambda_1 \) is fixed at 0.1. For inference, the regression model \( h\varphi \) is trained using ground-truth poverty indicators, employing a random forest regressor with 50 estimator trees. For the Pearson correlation maximization loss, the logarithm of nightlight intensity and average building floor area was used.

\paragraph{Baseline details.} We implemented several existing satellite representation learning models as baselines: (1) \textsf{Nightlight Proxy} trains the encoder using nightlight intensity as a proxy and, like our framework, employs Pearson maximization loss; (2) \textsf{Tile2Vec} is an unsupervised model that applies triplet loss, using nearby images as positive samples and distant images as negative samples to learn satellite image representations. Following the original implementation, the L2 norm is used as the distance metric, with a margin set to 0.5; (3) \textsf{READ} is a weakly supervised approach that leverages a small human-labeled dataset categorized into urban, rural, and uninhabited areas. Labels were assigned by four experts, and the resulting soft labels were used for training; (4) \textsf{PG-SimCLR} applies contrastive learning by generating positive samples based on POI proximity. The closest vector in terms of the L2 distance of the distance vector $\textbf{d}$ was utilized for pairing positive samples; and (5) \textsf{UrbanCLIP}, a multimodal learning model that integrates satellite images with text descriptions from a Vision-Language Model (VLM). Text descriptions generated by the satellite imagery VLM, TeoChat, were utilized in the training process~\cite{irvin2024teochat}.

\begin{table}[t!]
    \centering
    \renewcommand{\arraystretch}{1.1} 
    \resizebox{0.8\linewidth}{!}{%
    \begin{tabular}{l|l}
        \toprule
        \textbf{Dimensions of Poverty} & \textbf{Indicators} \\
        \midrule
        \multirow{2}{*}{Health} & Child Mortality \\
                                 & Nutrition \\
        \midrule
        \multirow{2}{*}{Education} & Years of Schooling \\
                                   & School Attendance \\
        \midrule
        \multirow{6}{*}{Standard of Living} & Cooking Fuel \\
                                            & Sanitation \\
                                            & Drinking Water \\
                                            & Electricity \\
                                            & Housing \\
                                            & Assets \\
        \bottomrule
    \end{tabular}
    }
    \caption{Multidimensional poverty and its corresponding indicators defined by the UN}
    \label{tab:poverty_indicators}
\end{table}

\subsection{Further Details of Result}
\begin{figure}[t!]
\centerline{
\includegraphics[width=0.7\linewidth]{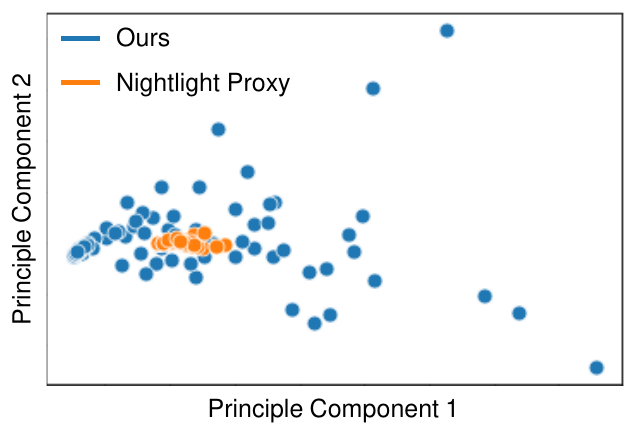}}
 
      \caption{Qualitative analysis of district representations using PCA. }
\label{fig:pca_score}
\end{figure}

\begin{figure*}[t!]
\centering
\includegraphics[width=0.8\textwidth]{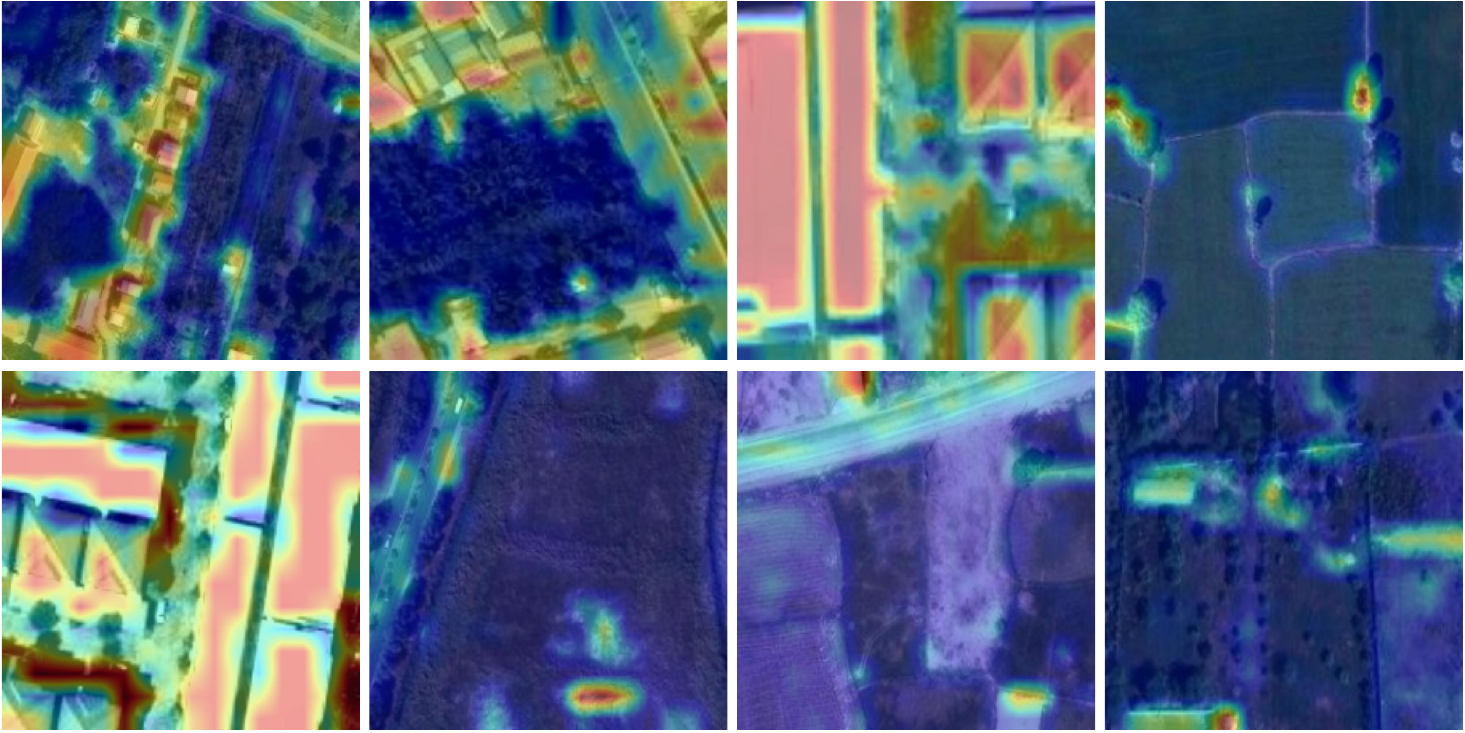}

\caption{Visualization of attention map for Backdoor Adjustment.}
\label{fig:attention}
\end{figure*}

Figure~\ref{fig:pca_score} reports the explainability scores for each PCA dimension, comparing our representation with the Nightlight Proxy representation. 
Figure~\ref{fig:attention} illustrates the attention map for backdoor adjustment. Instead of encompassing the entire image, the nightlight data specifically highlights regions associated with poverty. Table \ref{tab:performance_comparison_detail} presents the results, including standard deviations, for various baseline models. 

\begin{table*}[t]
    \centering
    \renewcommand{\arraystretch}{1.2} 
    \setlength{\tabcolsep}{6pt} 
    
    \begin{tabular}{l c c c}
        \toprule
        KHM & Pearson & Spearman  & $R^2$  \\
        \midrule
        Nightlight Proxy & 0.6666 $\pm$ 0.0323 & 0.6655 $\pm$ 0.0396 & 0.4346 $\pm$ 0.0439 \\
        Tile2Vec & 0.6552 $\pm$ 0.0294 & 0.6743 $\pm$ 0.0407 & 0.4170 $\pm$ 0.0382 \\
        READ & 0.6770 $\pm$ 0.0330 & 0.6748 $\pm$ 0.0409 & 0.4401 $\pm$ 0.0439 \\
        PG-SimCLR & 0.6750 $\pm$ 0.0266 & 0.6765 $\pm$ 0.0355 & 0.4454 $\pm$ 0.0378 \\
        UrbanCLIP & 0.6616 $\pm$ 0.0283 & 0.6529 $\pm$ 0.0326 & 0.4292 $\pm$ 0.0309 \\
        \midrule
        Ours & 0.7190 $\pm$ 0.0278 & 0.7437 $\pm$ 0.0323 & 0.4886 $\pm$ 0.0395 \\
        \bottomrule
    \end{tabular}
    
    \vspace{5mm}
    
    \begin{tabular}{l c c c}
        \toprule
        BGD & Pearson & Spearman  & $R^2$  \\
        \midrule
        Nightlight Proxy & 0.6315 $\pm$ 0.0433 & 0.5808 $\pm$ 0.0830 & 0.2972 $\pm$ 0.0406 \\
        Tile2Vec & 0.4653 $\pm$ 0.0600 & 0.3729 $\pm$ 0.0979 & 0.1883 $\pm$ 0.0513 \\
        READ & 0.4770 $\pm$ 0.0538 & 0.3594 $\pm$ 0.0972 & 0.2075 $\pm$ 0.0388 \\
        PG-SimCLR & 0.6341 $\pm$ 0.0475 & 0.5598 $\pm$ 0.0857 & 0.2961 $\pm$ 0.0388 \\
        UrbanCLIP & 0.6294 $\pm$ 0.0510 & 0.5240 $\pm$ 0.0888 & 0.2895 $\pm$ 0.0428 \\
        \midrule
        Ours & 0.6702 $\pm$ 0.0492 & 0.5924 $\pm$ 0.0753 & 0.3239 $\pm$ 0.0507 \\
        \bottomrule
    \end{tabular}
    
    \vspace{5mm}
    
    \begin{tabular}{l c c c}
        \toprule
        ZAF & Pearson & Spearman  & $R^2$  \\
        \midrule
        Nightlight Proxy & 0.5888 $\pm$ 0.1362 & 0.6541 $\pm$ 0.1647 & 0.5887 $\pm$ 0.0298 \\
        Tile2Vec & 0.5101 $\pm$ 0.1403 & 0.6200 $\pm$ 0.1335 & 0.4938 $\pm$ 0.0387 \\
        READ & 0.5419 $\pm$ 0.1499 & 0.6385 $\pm$ 0.1405 & 0.5257 $\pm$ 0.0416 \\
        PG-SimCLR & 0.5794 $\pm$ 0.1351 & 0.6465 $\pm$ 0.1651 & 0.5882 $\pm$ 0.0302 \\
        UrbanCLIP & 0.5465 $\pm$ 0.1440 & 0.6274 $\pm$ 0.1467 & 0.5352 $\pm$ 0.0390 \\
        \midrule
        Ours & 0.6088 $\pm$ 0.1314 & 0.6669 $\pm$ 0.1606 & 0.6083 $\pm$ 0.0299 \\
        \bottomrule
    \end{tabular}
    \caption{Evaluation results with standard deviation across different baselines, assessed on three cities in LMICs (KHM: Cambodia, BGD: Bangladesh, ZAF: South Africa).}
    \label{tab:performance_comparison_detail}
\end{table*}

\end{document}